\documentstyle[11pt,epsfig]{article}
\newcounter{subequation}[equation]

\makeatletter 
 
\def\thesubequation{\theequation\@alph\c@subequation}
\def\@subeqnnum{{\rm (\thesubequation)}}
\def\slabel#1{\@bsphack\if@filesw {\let\thepage\relax
   \xdef\@gtempa{\write\@auxout{\string
      \newlabel{#1}{{\thesubequation}{\thepage}}}}}\@gtempa
   \if@nobreak \ifvmode\nobreak\fi\fi\fi\@esphack}
\def\subeqnarray{\stepcounter{equation}
\let\@currentlabel=\theequation\global\c@subequation\@ne
\global\@eqnswtrue
\global\@eqcnt\z@\tabskip\@centering\let\\=\@subeqncr
$$\halign to \displaywidth\bgroup\@eqnsel\hskip\@centering
  $\displaystyle\tabskip\z@{##}$&\global\@eqcnt\@ne
  \hskip 2\arraycolsep \hfil${##}$\hfil
  &\global\@eqcnt\tw@ \hskip 2\arraycolsep
  $\displaystyle\tabskip\z@{##}$\hfil
   \tabskip\@centering&\llap{##}\tabskip\z@\cr}
\def\endsubeqnarray{\@@subeqncr\egroup
                     $$\global\@ignoretrue}
\def\@subeqncr{{\ifnum0=`}\fi\@ifstar{\global\@eqpen\@M
    \@ysubeqncr}{\global\@eqpen\interdisplaylinepenalty \@ysubeqncr}}
\def\@ysubeqncr{\@ifnextchar [{\@xsubeqncr}{\@xsubeqncr[\z@]}}
\def\@xsubeqncr[#1]{\ifnum0=`{\fi}\@@subeqncr
   \noalign{\penalty\@eqpen\vskip\jot\vskip #1\relax}}
\def\@@subeqncr{\let\@tempa\relax
    \ifcase\@eqcnt \def\@tempa{& & &}\or \def\@tempa{& &}
      \else \def\@tempa{&}\fi
     \@tempa \if@eqnsw\@subeqnnum\refstepcounter{subequation}\fi
     \global\@eqnswtrue\global\@eqcnt\z@\cr}
\let\@ssubeqncr=\@subeqncr
\@namedef{subeqnarray*}{\def\@subeqncr{\nonumber\@ssubeqncr}\subeqnarray}
\@namedef{endsubeqnarray*}{\global\advance\c@equation\m@ne%
                           \nonumber\endsubeqnarray}

\makeatletter
\@addtoreset{equation}{section}
\makeatother
\renewcommand{\theequation}{\thesection.\arabic{equation}}

\def\dalemb#1#2{{\vbox{\hrule height .#2pt
        \hbox{\vrule width.#2pt height#1pt \kern#1pt
                \vrule width.#2pt}
        \hrule height.#2pt}}}
\def\square{\mathord{\dalemb{6.8}{7}\hbox{\hskip1pt}}}

    \let\e=\epsilon
  \let\q=\theta  
  \let\n=\nu

 \def\bd{\begin{document}} \def\ed{\end{document}}
\def\ds{\documentstyle} \let\fr=\frac \let\bl=\bigl \let\br=\bigr
\let\Br=\Bigr \let\Bl=\Bigl 
\let\bm=\bibitem
\let\na=\nabla
\let\pa=\partial \let\ov=\overline
\def\ie{{\it i.e.\ }} 
\newcommand{\be}{\begin{equation}} 
\newcommand{\ee}{\end{equation}} 
\def\ba{\begin{array}}
\def\ea{\end{array}}
\def\ft#1#2{{\textstyle{{\scriptstyle #1}\over {\scriptstyle #2}}}}
\def\fft#1#2{{#1 \over #2}}
\def\del{\partial}
\def\sst#1{{\scriptscriptstyle #1}}
\def\oneone{\rlap 1\mkern4mu{\rm l}}
\def\e7{E_{7(+7)}}
\def\td{\tilde}
\def\wtd{\widetilde}
\def\im{{\rm i}}
\def\bog{Bogomol'nyi\ }
\def\q{{\tilde q}}
\def\hast{{\hat\ast}}
\def\0{{\sst{(0)}}}
\def\1{{\sst{(1)}}}
\def\2{{\sst{(2)}}}
\def\3{{\sst{(3)}}}
\def\4{{\sst{(4)}}}
\def\5{{\sst{(5)}}}
\def\6{{\sst{(6)}}}
\def\7{{\sst{(7)}}}
\def\8{{\sst{(8)}}}
\def\n{{\sst{(n)}}}
\def\oo{{\"o}}
\def\hA{\hat{\cal A}}
\def\ns{{\sst {\rm NS}}}
\def\rr{{\sst {\rm RR}}}
\def\tH{{\widetilde H}}
\def\tB{{\widetilde B}}
\def\cA{{\cal A}}
\def\cF{{\cal F}}
\def\tF{{\wtd F}}
\def\Z{\rlap{\sf Z}\mkern3mu{\sf Z}}
\def\ep{{\epsilon}}
\def\IIA{{\rm IIA}}
\def\IIB{{\rm IIB}}
\def\ads{{\rm AdS}}
\def\R{\rlap{\rm I}\mkern3mu{\rm R}}
\def\mapright#1{\smash{\mathop{-\!\!\!-\!\!\!-\!\!\!-\!\!\!-\!\!\!
             \longrightarrow}\limits^{#1}}}

\def\Ei{{\hbox{Ei}}}
\def\Ci{{\hbox{Ci}}}
\def\Si{{\hbox{Si}}}

\newcommand{\ho}[1]{$\, ^{#1}$}
\newcommand{\hoch}[1]{$\, ^{#1}$}
\newcommand{\bea}{\begin{eqnarray}} 
\newcommand{\eea}{\end{eqnarray}} 
\newcommand{\ra}{\rightarrow}
\newcommand{\lra}{\longrightarrow}
\newcommand{\Lra}{\Leftrightarrow}
\newcommand{\aap}{\alpha^\prime}
\newcommand{\bp}{\tilde \beta^\prime}
\newcommand{\tr}{{\rm tr} }
\newcommand{\Tr}{{\rm Tr} } 
\newcommand{\NP}{Nucl. Phys. }
\newcommand{\tamphys}{\it Center for Theoretical Physics,
Texas A\&M University, College Station, TX 77843}

\newcommand{\upenn}{\it Department of Physics and Astronomy,\\ University
of Pennsylvania, Philadelphia, PA 19104}

\newcommand{\brussels}{\it Physique Th\'eorique et Math\'ematique, 
Universit\'e Libre de Bruxelles,\\ Campus Plaine C.P. 231, B-1050
Bruxelles, Belgium} 

\newcommand{\auth}{}

\begin{document}
\begin{flushright}
ULB-TH/02-02\\
February  2002\\
\hfill{\bf hep-th/0202029}\\
\end{flushright}


\begin{center}

{\large {\bf Rotational Graviton Modes in the Brane World}}

\vspace{20pt}

\auth{J.F. V\'{a}zquez-Poritz}

\vspace{10pt}
\brussels\\
\vspace{10pt}

\vspace{30pt}

\underline{ABSTRACT}
\end{center}
For a brane world embedded in various ten or eleven-dimensional geometries, we
calculate the corrections to the four-dimensional gravitational potential due to
graviton modes propagating in the extra dimensions, including those rotating
around compact directions. Due to additional "warp" factors, these
rotation modes may have as significant an effect as the s-wave modes which
propagate in the large or infinite extra dimension.

{\vfill\leftline{}\vfill
\vskip 10pt \footnoterule {\footnotesize \hoch{1} This work is supported
in part by the Francqui Foundation (Belgium), the Actions de Recherche
Concert{\'e}es of the Direction de la Recherche Scientifique - Communaut\'e
Francaise de Belgique, IISN-Belgium (convention 4.4505.86).
 
\vskip  -12pt} \vskip   14pt
}

\pagebreak
\setcounter{page}{1}


\section{Introduction}

It has been proposed that there may exist extra dimensions which are almost the
size of a millimeter \cite{hamed1, hamed2}. In this case, the gravitational
potential would go as $1/r^{1+n}$ for sub-millimeter measurements, where
$n$ is the number of extra millimeter dimensions. The four-dimensional potential
is recovered at larger scales. If the four-dimensional metric is multiplied by a
"warp" factor which is a rapidly changing function of an extra dimension, then
four-dimensional gravity is recovered at low-energy scales even for an infinite
extra dimension \cite{randall1,randall2}. This is due to a massless graviton state
that is bound to a 3-brane embedded in the higher-dimensional space. Large-energy
modifications to the four-dimensional potential would result from the Kaluza-Klein
modes propagating in the extra dimension.

It is possible that nature has chosen some extra dimensions to be infinite and
others to be compact, especially if our spacetime is embedded within M-theory.
For example, five-dimensional domain walls which localize gravity may arise from a 
sphere reduction from ten or eleven dimensions, as the near-horizon of extremal
$p$-brane configurations \cite{cvetic}\footnote{It has been found that a
non-extremal brane origin yields a localized graviton of nonzero
mass \cite{jvp2}.}. In particular, it was found that such domain
walls can be derived from a D3, D4 (or M5) and D5-brane, as well as two
M5-branes intersecting over a 3-brane\footnote{Other $p$-brane origins
are possible with the inclusion of a naked singularity \cite{youm}.}. 

In embedding five-dimensional brane world scenarios in higher dimensions, one
generally only considers the s-wave Kaluza-Klein graviton mode. That is, as part
of the dimensional reduction ansatz, one discards modes moving about the compact
directions. It would appear to be a reasonable assumption that, unless the compact
dimensions are quite large, the effects of such rotation modes on the corrections
to the effective four-dimensional gravitation would be over-shadowed by the
s-wave Kaluza-Klein modes propagating along the infinite extra dimension. However,
it once appeared to be reasonable to assume that a theory with more than 1+3 large
dimensions did not agree with our direct observations. 

It was found that the "warp" factor in certain five-dimensional spaces is such that 
the effects of gravitational modes propagating in the extra dimension is dampened,
yielding four-dimensional gravity at low energies. Embedding the brane world in
ten or eleven dimensions brings the possibility of additional "warp" factors.
In particular, the effects of modes rotating about compact dimensions could be
amplified if the corresponding metric elements are multiplied by a "warp" factor
which increases rapidly along the infinite  extra dimension. Depending
on the geometry of the higher-dimensional origin, rotational Kaluza-Klein modes
may actually be as significant as the s-wave Kaluza-Klein
modes\footnote{Throughout this paper, we assume that all compact dimensions are
the same size, fixed by the length scale 1/$k$ in the "warp" factors.}. 

This paper is organized as follows. In section $p-1$ we analyse how the
rotational graviton modes modify the four-dimensional gravitational
potential for a brane world originating from a D$p$-brane, where
$p=3,4,5$. In these cases, we find that the rotational modes do not
contribute significantly to the modification of the gravitational potential
relative to the corrections due to the s-wave Kaluza-Klein modes. It is
interesting to note that, in some cases, there are rotational modes that are
localized on the brane world. In section 5, we are pleasantly surprised to find
that the metric for the near-horizon of the M5/M5 system contains a "warp" factor
which amplifies the effect of the rotational graviton modes. In this case, the
corrective contributions from the s-wave Kaluza-Klein graviton modes and the
rotational Kaluza-Klein graviton modes are of the same magnitude. We offer
Conclusions in section 6.

\section{D3-brane origin}

The D3-brane metric is
\be
ds_{10}^2=H^{-1/2}(-dt^2+dx_i^2)+H^{1/2}(dr^2+r^2
d\Omega_{5}^2),
\ee
where
\be
H=1+\frac{R^4}{r^4},
\ee
and $i=1,2,3$. In the near-horizon limit $r<<R$, the above metric
is that of $AdS_5 \times S^5$. This can be expressed as
\be
ds_{10}^2=(1+k|z|)^{-2}(-dt^2+dx_i^2+dz^2)+\frac{1}{k^2} d\Omega_5^2,
\label{metric}
\ee
where 
\be
\frac{r}{R}=(1+k|z|)^{-1},
\ee
and $R=k^{-1}$. Note that the absolute value in the above coordinate
transformations imposes $Z_2$ symmetry, which has no known
source in ten dimensions\footnote{Recently, it has been found that
resolved branes on an Eguchi-Hanson instanton dimensionally reduce to
five-dimensional domain walls which localize gravity, without the need
of an extra source term\cite{jvp}.}.

The equation of motion for a graviton fluctuation is that of a
minimally-coupled scalar:
\be
\partial_M \sqrt{-g}g^{MN}\partial_N \Phi=0.
\ee
We take $\Phi=P_{\ell}(S^5)(1+k|z|)^{3/2}\psi(z)M(t,x_i)$, where
$\square_{(4)}M=m^2 M$ and $\square_{(4)}$ is the Laplacian on
$t,x_i$. $P_{\ell}$ are the Legendre polynomials of $S^5$. For graviton modes
on the metric (\ref{metric}) we obtain
\be
-\partial_z^2 \psi+\Big[ \frac{(\ell+3/2)(\ell+5/2)k^2}{(1+k|z|)^2}
-3k\delta(z)\Big] \psi=m^2 \psi,
\ee
where $\ell$ is an integer parametrizing the rotational dynamics about
$S^5$. For the massless modes, the solution is
\be
\psi=(1+k|z|)^{-\ell-3/2}+\frac{\ell}{\ell+4}(1+k|z|)^{\ell+5/2}.
\ee
For $\ell=0$, this wave function is normalizable and there is a
corresponding localized graviton. However, for $\ell >0$ the wave
function diverges for large $z$, and thus there is no permissible
massless solution. The massive wave functions are given by
$$
\psi_m=N_m(1+k|z|)^{1/2}\Big[ \Big( Y_{1+\ell}\Big(
\frac{m}{k}\Big) -\ell 
\frac{k}{m}Y_{2+\ell}\Big( 
\frac{m}{k}\Big) \Big) J_{2+\ell}\Big( \frac{m}{k}(1+k|z|)\Big) + $$
\be
\Big( \ell \frac{k}{m}J_{2+\ell}\Big( 
\frac{m}{k}\Big) -J_{1+\ell}\Big( \frac{m}{k}\Big)   
\Big) Y_{2+\ell}\Big( \frac{m}{k}(1+k|z|)\Big) \Big],
\ee
where
\be
N_m=\frac{1}{\sqrt{2k}}\Big[ \Big( Y_{1+\ell}\Big(
\frac{m}{k}\Big) -\ell \frac{k}{m}Y_{2+\ell}\Big( \frac{m}{k}\Big) 
\Big)^2+\Big( \ell \frac{k}{m} J_{2+\ell}\Big(
\frac{m}{k}\Big) -J_{1+\ell}\Big( \frac{m}{k}\Big) \Big)^2 \Big]^{-1/2}.
\ee

The Newtonian gravitational potential between masses $M_1$ and $M_2$ can be
estimated by
\be
U(r)\sim 
\sum_{\ell=0}^{\infty}\Big( \frac{G_4 M_1 M_2}{r}{\rm e}^{-m_{\ell}r}
+\frac{G_5M_1 M_2}{r}\int_{m_0^2}^{\infty}d(m^2) |\psi_m(z=0,\ell)|^2 {\rm
e}^{-mr}\Big) ,\label{pot}
\ee
where $m_{\ell}$ is the mass of a bound state and $m_0$ is the minimum mass
for non-localized states, each for a given $\ell$. In the
present case, there are no massive bound states or mass gaps due to $\ell
>0$ states, so that $m_0=m_{\ell}=0$. The four and five-dimensional Newton 
constants are related by $G_4=kG_5$. This yields
\be
U(r)\sim \frac{G_4 M_1 M_2}{r}\Big( \Big[ 1+\frac{c_0}{(kr)^2}+...\Big] 
+\sum_{\ell=1}^{\infty}\Big[ \frac{c_{\ell}}{(kr)^{6+2\ell}}+... \Big]
\Big),
\ee
where $c_0$ and $c_{\ell}$ are constants of order one. The $\ell=0$
terms were found in \cite{randall1,cvetic}. The terms in the second
pair of square brackets are the dominant corrections from the
rotational graviton modes ($\ell >0$). As can be seen, the
dominant contribution from the $\ell >0$ modes is smaller than that of
the s-wave Kaluza-Klein modes by the rather small factor $1/(kr)^6$.

\section{D4-brane origin}

The metric for the D4-brane is
\be
ds_{10}^2=H^{-3/8}(-dt^2+dx_i^2)+H^{5/8}(dr^2+r^2 d\Omega_4^2),
\ee
where
\be
H=1+\frac{R^3}{r^3},
\ee
and $i=1,..,4$. In the near-horizon limit $r<<R$, the above metric
can be expressed as
\be
ds_{10}^2=(1+k|z|)^{-9/4}(-dt^2+dx_i^2+dz^2)+(1+k|z|)^{-1/4}\frac{1}{k^2}
d\Omega_5^2, \label{metric2}
\ee
where 
\be
\frac{r}{R}=(1+k|z|)^{-2},
\ee
The analysis is similar to the case of the D3-brane. An additional
element is that the worldvolume direction $x_4$ is taken to be
compact, in order for the worldvolume of the braneworld to have 1+3
dimensions. Rotational dynamics around $x_4$ are parametrized by the
integer $n$. For simplicity, we will assume that all compact
dimensions are of the same size. From (\ref{pot}) we find that 
\be 
U(r)\sim \frac{G_4 M_1 M_2}{r}\sum_{n=0}^{\infty}{\rm e}^{-nkr}
\Big( \Big[
1+\frac{b_0}{(kr)^4}+... \Big] 
+\sum_{\ell=1}^{\infty}\Big[
\frac{b_{\ell}}{(kr)^{2\sqrt{\ell^2+3\ell+9}+2}}+... \Big] \Big),
\label{D4pot}
\ee
where $b_0$ and $b_{\ell}$ are constants of order one. The $\ell=0, M=0$
terms were found in \cite{cvetic}. The terms in the second pair of square
brackets are the dominant corrections for the rotational graviton
modes ($\ell >0$). As can be seen, the dominant contribution from the $\ell
>0$ modes is smaller than that of the s-wave Kaluza-Klein modes by a rather
small factor of approximately $1/(kr)^5$. An additional element here, as
opposed to the D3-brane origin, is the presence of massive bound states and
additional Kaluza-Klein states, both due to the $n>0$ modes. These 
contribute Yukawa-like terms to the gravitational potential. In the case of
an $M5$-brane, the result is the same as above, except that now there is the
possibility of modes rotating around two compact worldvolume dimensions, so
that the $n$ in (\ref{D4pot}) is replaced by $\sqrt{n_1^2+n_2^2}$.

\section{D5-brane origin}

The metric for the D5-brane is
\be
ds_{10}^2=H^{-1/4}(-dt^2+dx_i^2)+H^{3/4}(dr^2+r^2 d\Omega_3^2),
\ee
where
\be
H=1+\frac{R^2}{r^2},
\ee
and $i=1,..,5$. In the near-horizon limit $r<<R$, the above metric
can be expressed as
\be
ds_{10}^2={\rm e}^{-k|z|/4}(-dt^2+dx_i^2+dz^2+\frac{1}{k^2}d\Omega_3^2),
\label{metric3}
\ee
where 
\be
\frac{r}{R}={\rm e}^{-k|z|/2},
\ee

In order to yield a braneworld with 1+3 worldvolume dimensions, we take
$x_4$ and $x_5$ to be compact. We take $\Phi={\rm
e}^{i(n_1x_4+n_2x_5)}P_{\ell}(S^3){\rm
e}^{k|z|/2}\psi(z)M(t,x_i)$. For graviton
modes on the metric (\ref{metric3}) we obtain
\be
-\partial_z^2 \psi+\Big[ \Big( \frac{1}{4}+\alpha^2\Big) k^2
-k\delta(z)\Big] \psi=m^2 \psi,\label{wave2}
\ee
where $\alpha^2 \equiv \ell (\ell +2)+n_1^2+n_2^2$. $\ell$, $n_1$
and $n_2$ are integers parametrizing the rotational dynamics around
$S^3$, $x_4$ and $x_5$ respectively. For the massless modes, the solution
is 
\be
\psi={\rm e}^{-\sqrt{\alpha^2+1/4}k|z|}+A {\rm e}^{\sqrt{\alpha^2+1/4} 
k|z|},\label{zero}
\ee
where
\be
A\equiv \frac{2\sqrt{\alpha^2+1/4}-1}{2\sqrt{\alpha^2+1/4}+1}.
\ee
For $\ell,n_1,n_2=0$, this wave function is normalizable and there is a
corresponding localized graviton. However, for all higher angular
momentum modes, the wave function diverges for large $z$ and thus there
is no permissible massless solution. Nonzero $\ell,n_1,n_2$
effectively decrease the mass term in the wave equation 
(\ref{wave2}). Thus, a higher angular momentum mode has a localized state
on the brane world of mass $m^2=(\ell (\ell+2)+n_1^2+n_2^2)k^2$. Also,
each mode has an equal mass gap between the bound state and a continuum
of Kaluza-Klein modes.

The massive wave functions are given by
$$
\psi_m=N_m (k\sin q|z|-2q\cos q|z|),
$$
\be
N_m=\frac{1}{2\sqrt{\pi q(m^2-\alpha ^2 k^2)}},
\ee
$$
q=\sqrt{m^2-k^2(1/4+\alpha^2)}.
$$
From (\ref{pot}) we find that
\be
U(r)\sim \frac{G_4 M_1 M_2}{r} \sum_{\ell,n_1,n_2=0}^{\infty}\Big[ {\rm
e}^{-\alpha kr}+
\frac{2}{\sqrt{\pi}}\Big(
1+4\alpha^2 \Big)^{3/4} {\rm e}^{-\sqrt{\alpha^2+1/4}kr}
\Big( \frac{1}{(kr)^{3/2}}+...\Big) \Big], \label{D5pot}
\ee
The $\ell,n_1,n_2=0$ terms were found in \cite{cvetic}. As can be seen,
the rotational graviton modes yield Kaluza-Klein potential terms whose
dominant correction is smaller than that of the s-wave Kaluza-Klein terms
by a relative Yukawa-like factor of approximately ${\rm
e}^{-kr}$. Notice that there are also contributions from rotational bound
states, whose dominant term is smaller than that of the s-wave
Kaluza-Klein modes by a relative Yukawa-like factor of ${\rm e}^{-kr/2}$.

\section{M5/M5 origin}

The metric for two M5-branes intersecting over 1+3 dimensions is
\be
ds_{11}^2=H_1^{-1/3}H_2^{-1/3}\Big( -dt^2+dx_i^2+H_1(dy_1^2+dy_2^2) 
+H_2(dy_3^2+dy_4^2)+H_1H_2(dr^2+r^2d\Omega_2^2)\Big),
\ee
where
\be
H_{1(2)}=1+\frac{R_{1(2)}}{r},
\ee
and $i=1,2,3$. For $R_1=R_2\equiv R$ and in the near-horizon limit
$r<<R$, the above metric can be expressed as
\be
ds_{11}^2={\rm e}^{-\frac{2}{3}k|z|}(-dt^2+dx_i^2+dz^2+
\frac{1}{k^2}d\Omega_2^2)+{\rm e}^{\frac{1}{3}k|z|}dy_j^2,
\label{metric4}
\ee
where 
\be
\frac{r}{R}={\rm e}^{-k|z|},
\ee
and $j=1,..,4$. We take $y_j$ to be compact. Notice that the metric elements
corresponding to the compact M5-brane worldvolume directions $y_j$ are multiplied
by a "warp" factor, which increases rapidly with $|z|$. It will be shown how this
"warp" factor amplifies the effect that the graviton modes rotating about $y_j$
have on the corrections to the four-dimensional gravitational potential. 

We take $\Phi={\rm e}^{i(n_j y_j)}P_{\ell}(S^2){\rm
e}^{k|z|/2}\psi(z)M(t,x_i)$, where we shall implicitly sum over $j$
hereafter. For graviton modes on the metric given by (\ref{metric4}), we
obtain
\be
-\partial_z^2 \psi+\Big[ (\ell+1/2)^2 k^2+n_j^2 k^2{\rm e}^{-k|z|}
-k\delta(z)\Big] \psi=m^2 \psi,
\ee
The solution is 
$$
\psi=N_m\Big[ \Big( (i\gamma+1)Y_{i\gamma }(2in)-2inY_{i\gamma-1}(2in) \Big) 
J_{i\gamma}\Big( 2in{\rm e}^{-k|z|/2}\Big) 
+
$$
\be
\Big( 2inJ_{i\gamma -1}(2in)-(i\gamma +1)J_{i\gamma }(2in) \Big) 
Y_{i\gamma }\Big( 2in{\rm e}^{-k|z|/2}
\Big) \Big],
\ee
where $\gamma \equiv \sqrt{4m^2/k^2-(2\ell+1)^2}$ and $n^2\equiv
n_j^2$. The higher $\ell$ modes  contribute massive bound states at
$m^2=\ell(\ell+1)k^2$ and equal mass gaps between the bound states and
continua of Kaluza-Klein modes. The $n>0$ modes only contribute the
latter. For $m > (\ell +2)k$, the Kaluza-Klein modes can be plane-wave
normalized with
$$
N_m=\frac{1}{2\sqrt{k}}\Big[ {\rm sinh}(\pi \gamma )
\Big| (i\gamma+1)Y_{i\gamma }(2in)-2inY_{i\gamma-1}(2in) \Big| ^2+
$$
\be
\frac{1}{{\rm sinh}(\pi \gamma)}
\Big| 2inJ_{i\gamma -1}(2in)-(i\gamma +1)J_{i\gamma }(2in) \Big| ^2
\Big] ^{-1/2}.
\ee
From (\ref{pot}) we find that
$$
U(r)\sim \frac{G_4 M_1 M_2}{r} \sum_{\ell =0}^{\infty}\Big[ {\rm
e}^{-\sqrt{\ell(\ell+1)}kr}+
\Big( \frac{2}{\sqrt{\pi}}(2\ell+1)^{3/2}+
$$
\be
\sqrt{2}\pi^{3/2}
(2\ell+1)\sum_{j=1}^{4}\sum_{n_j=1}^{\infty}n^2\Big|
\frac{J_{-1}(2in) Y_0(2in) 
-J_0(2in)Y_{-1}(2in)}{2in Y_{-1}(2in)-J_0(2in)}
\Big|^2 \Big) {\rm e}^{-(\ell+1/2)kr}
(\frac{1}{(kr)^{3/2}}+...) \Big], \label{M5M5pot}
\ee
For the D5 and M5/M5 origins, the contributions to the gravitational potential
are the same for the $\ell,n=0$ modes \cite{cvetic} but different for the
rotational modes. As can be seen, the $\ell >0$ modes yield potential terms that
are sub-leading to the s-wave terms by a relative Yukawa-like factor, as
in the case of the D5-brane origin. Notice there are contributions from
massive bound states as well as additional Kaluza-Klein modes. 

The $n>0$ Kaluza-Klein modes contribute to the Newtonian potential on the
same order in $kr$ as the s-wave Kaluza-Klein modes! In particular, if we
consider the $\ell=0$ modes, we have
\be
U(r)\sim \frac{G_4 M_1 M_2}{r} \Big[ 1+\frac{2}{\sqrt{\pi}}\beta
{\rm e}^{-kr/2} (\frac{1}{(kr)^{3/2}}+...) \Big], \label{finalU}
\ee
where the factor $\beta=1.2$ is due to the presence of the $n=1$
modes. Of course this factor is larger if the length scale of the $y_j$
directions is larger than $1/k$. Modes with $n>1$ do not make such a significant
contribution.

\section{Conclusions}

We have found that, for brane worlds originating from the near-horizon of a
D$p$-brane where $p=3,4,5$, the dominant correction of the four-dimensional
gravitational potential arises from s-wave Kaluza-Klein modes propagating in the
infinite extra dimension. The effect of Kaluza-Klein modes rotating about compact
dimensions is sub-leading to that of the s-wave Kaluza-Klein modes, regardless of
whether the former states are bound to the brane world (with Yukawa-like
factors) or are propagating in the infinite extra dimension. This result is
perhaps to be expected.

On the other hand, for a brane world scenario embedded in the near-horizon region
of an M5/M5 intersection, we have found that there is a "warp" factor
which amplifies the effect of graviton modes rotating about compact directions
that lie in the worldvolumes of the M5-branes. In this case, in fact, the
modification of the four-dimensional gravitational potential due to these
rotational modes is on the same order as that due to the s-wave Kaluza-Klein
states. This can be seen explicitly by the presence of $\beta$ in (\ref{finalU}).
It is rather surprising that, the effects of modes around compact
directions can be considerably amplified by "warp" factors which depend on an
extra large or infinite dimension. More complicated geometries and
compactification schemes could yield additional amplifications in the
gravitational sector.

\section*{Acknowledgments}

We are grateful to Malcolm Fairbairn and Fernando Quevedo for useful
discussions.

\end{document}